\documentclass{article}
\usepackage{oldgerm}
\usepackage{yfonts}
\usepackage{euler}
\usepackage{graphics}
\usepackage{bm}
\usepackage{graphicx}
\usepackage{amsmath}
\usepackage{amssymb}
\usepackage{amstext}
\usepackage{amscd}
\usepackage{amsfonts}
\usepackage{appendix}
\usepackage{wasysym}
\usepackage{epstopdf}
\usepackage{color}
\usepackage{hyperref}
\DeclareMathAlphabet{\EuFrak}{U}{euf}{m}{n}
\DeclareMathAlphabet{\EuScript}{U}{eus}{m}{n}

\newcommand{\nd}{\noindent}

\newcommand{\be}{\begin{equation}}
\newcommand{\ee}{\end{equation}}
\newcommand{\ben}{\begin{eqnarray}}
\newcommand{\een}{\end{eqnarray}}

\title{{\bf Supercoherent states of the open NS world sheet superstring.}}
\author{{Mir Hameeda$^{1,2}$, Mario C. Rocca$^{3,4,5}$ }  \\
\small{$^1$ Department of Physics, S.P. Collage, Srinagar, Kashmir, 190001 India}\\
\small{$^2$ Inter University Centre for Astronomy and Astrophysics , Pune India}\\
\small{$^3$ Departamento de F\'{\i}sica,
Universidad Nacional de La Plata. Argentina}\\
\small{$^4$ Departamento de Matem\'{a}tica,
Universidad Nacional de La Plata. Argentina}\\
\small{$^5$ Consejo Nacional de Investigaciones Cient\'{\i}ficas
y Tecnol\'{o}gicas}\\
\small{(IFLP-CCT-CONICET)-C. C. 727, 1900 La Plata -
Argentina}}
\date{\today}

\begin{document}

\maketitle

\vspace{-5mm}

\begin{abstract}
\nd
The supercoherent states of the RNS string are constructed using the covariant quantization and analogously the light cone quantization formalisms.  Keeping intact the original definition of coherent states of harmonic oscillators, we extend the bosonic annihalation operator into the superspace by inclusion of fermionic contribution to oscillator modes thus construct the supercoherent states with supersymetric harmonic oscillator. We analyse the statistics of these states by explicitly calculating the Mandel parameter and obtained interesting results about the nature of distribution of the states.\\
\nd
{\bf KEYWORDS}: Coherent states; supercoherent states; harmonic oscillator; bosonic string; fermionic string.\\

\end{abstract}

\newpage

\tableofcontents

\newpage

\renewcommand{\theequation}{\arabic{section}.\arabic{equation}}

\section{Introduction}
With all its beauties intact, a few shortcomings, especially the absence of fermions and the presence of tachyons, made physists uneasy and prompted them to extend the bosonic string theory to the wider aspects of supersymmetric string theory, by welcoming the inclusion of fermionic partners with promising outcomes. 
 Supersymmetric string theory shortly called as Superstring theory is actually the version of string theory that accounts for both fermions and bosons and incorporates the concept of supersymmetry in understanding the deeper insights of the subject. Supersymmetry involves symmetry and a mathematical transformation between bosons and fermions. Keeping in view that that the free fermions contain states that transform as spinors under the associated orthogonal symmetry, it became easy to to add free fermions on the world-sheet of the string in order to obtain states that transform as spinors. \cite{gsw}
Rediscovered and studied by many \cite{fe},\cite{tw}, SUSY, the popular abbreviation for supersymmetry, was originally introduced by Gelfand and Likhtman in 1971 \cite{gl}. The other workers who should be equally credited are Ramond,  Neveu, and Schwartz. Witten(1982) \cite{gsw} extensively developed the supersymmetric quantum mechanics. Supersymmetry has played a very crucial role in understanding the theoretical aspects of quantum field theories and the gravity theories. The concept of supersymmetry is being explored to understand the finer details of the string theories. \cite{aa}.

Supersymmetry gives rise to the concept of supercoherent states with supersymetric harmonic oscillator a combo of fermionic and bosonic harmonic oscillators, as the most preferable foundation for the construction of supercoherent states. The basic definition of supercoherent state is as the eigenstates of a super-annihilation operator. Super-annihilation operator in infact is only an extension of the bosonic operator into superspace.

Even and odd coherent states of supersymmetric harmonic oscillators and their nonclassical properties have been studied
\cite{dsf}.  
In string theory the supercoherent states and its allied properties is yet in its budding bloom and needs to be given a deep attention.
This paper is especially dedicated to understand the concept supercoherent states based on the quantum mechanical analogue of harmonic oscillator and the extension of the bosonic coherent states. This is a bold and novel try to use the concept of superpotential \cite{aa} and the differential representation of the super-annihilator operator to analytically derive the supercoherent states. 

Supersymmetry is incorporated into string theory by two approaches, the Ramond-Neveu-Schwarz (RNS) formalism and the Green-Schwarz (GS) formalism \cite{bbs}. In this paper we will work with the RNS string spectrum which inspite of surviving the super virasoro constrains shows some problems especially in the NS sector where the ground state is a tachyon and some inconsistency in the symmetry of fermions and bosons in the zero modes. But the essential requirement of consistent interacting theory is the unbroken supersymmetry which in our case is maintained except the fermionic zero modes. Thus the GSO projection procedure introduced by Gliozzi, Scherk and Olive will be followed to get rid of the supersymmetry breaking.The GSO projection leaves an equal number of bosons and fermions
at each mass level thus supports that the theory has a spacetime supersymmetry. The statistical properties in the form of Mandal parameter \cite{man} will be a beautiful extension of this paper.

\section{Quantization of the superstring in Ramond Neveu Schwarz (RNS) formalism}
The RNS formalism involves the pairing of bosonic fields $X^\mu(\sigma,\tau)$ of the two dimensional world sheet with the fermionic fields $\Psi^\mu(\sigma,\tau)$, the two component spinors of the world sheet and vectors under Lorentz transformation of the D-dimensional space time. Being spinors, these fields follow the anticommutation relations besides keeping consistency with the spin and statistics in $D=10$ dimensions. The resulting action from pairing of the two fields, invariant under supersymmetry transformations is expressed as
\begin{equation}
\label{eq2.1}
S=-\frac{1}{2\pi}\int d^2\sigma\left(\partial_\alpha X_\mu\partial^\alpha X^\mu+\bar\Psi^\mu\rho^\alpha\partial_\alpha\Psi_\mu\right)
\end{equation}
where $\rho^\alpha$, with $\alpha=0,1$, are the two dimensional Dirac matrices.

The quantizations schemes of covariant and light cone gauge as followed for bosonic string theory is adopted here with an additional requirement of supersymmetry as demanded by inclusion of fermions in the theory. The two component spinor is given by
\begin{equation}
\label{eq2.2}
\Psi^\mu=\begin{pmatrix}
\Psi_-^\mu\\
\Psi_+^\mu
\end{pmatrix}
\end{equation}
The detailed analysis of boundary conditions and mode expansion has been explained in \cite{bbs}, \cite{gsw}. 
In the case of Ramond (R) boundary conditions $\Psi_+^\mu(\pi,\tau)=\Psi_-^\mu(\pi,\tau)$, the mode expansion of the Dirac equation are:
\begin{equation}
\label{eq2.3}
\Psi_-^\mu(\sigma,\tau)=\frac{1}{\sqrt{2}}\sum_n d_n^\mu e^{-in(\tau-\sigma)}
\end{equation}
\begin{equation}
\label{eq2.4}
\Psi_+^\mu(\sigma,\tau)=\frac{1}{\sqrt{2}}\sum_n d_n^\mu e^{-in(\tau+\sigma)}
\end{equation}
where the sum runs over all integers $n$.
In case of Neveu-Schwarz (NS) boundary conditions $\Psi_+^\mu(\pi,\tau)=-\Psi_-^\mu(\pi,\tau)$, the mode expansions become
\begin{equation}
\label{eq2.5}
\Psi_-^\mu(\sigma,\tau)=\frac{1}{\sqrt{2}}\sum\limits_{r\in\mathbb{Z}+\frac {1} {2}} b_r^\mu e^{-ir(\tau-\sigma)}
\end{equation}
\begin{equation}
\label{eq2.6}
\Psi_+^\mu(\sigma,\tau)=\frac{1}{\sqrt{2}}\sum\limits_{r\in\mathbb{Z }+\frac {1} {2}}b_n^\mu e^{-ir(\tau+\sigma)}
\end{equation}
where the sum runs over half-integer modes $r$.\\

In the covariant gauge the dynamics of bosonic and fermionic coordinates are given respectively, by free two dimensional Klein Gordon equation and a free Dirac equation augmented by certain constraints. Thus the quantization of the coordinates is similar to that of the two dimensional field theory. The commutations relations of bosonic fields and the corresponding Fourier coefficients are expressed as

\begin{equation}
\label{eq2.7}
[\dot{X}^\mu(\sigma,\tau),X^\nu(\sigma^{'},\tau)]=-i\pi\delta(\sigma-\sigma^{'})\eta^{\mu\nu}
\end{equation}
\begin{equation}
\label{eq2.8}
[\alpha_m^\mu,\alpha_n^\nu]= m\delta_{m+n}\eta^{\mu\nu}
\end{equation}
Where the $\alpha_m^\mu$ are coefficients in an open or closed string mode expansion. For the closed string there is again a second set of modes $\tilde\alpha_n$. \cite{bbs}

The quantization of the fermionic leads to the canonical anticommutation relations as:
\begin{equation}
\label{eq2.9}
\{\Psi_A^\mu(\sigma,\tau),\Psi_B^\nu(\sigma^{'},\tau)\}=\pi\delta(\sigma-\sigma^{'})\eta^{\mu\nu}\delta_{AB}
\end{equation}
and the corresponding half integrally mode and integrally mode oscillators respectively satisfy
\begin{equation}
\label{eq2.10}
\{b_r^\mu, b_s^\nu\}=\eta^{\mu\nu}\delta_{r+s}
\end{equation}
\begin{equation}
\label{eq2.11}
\{d_m^\mu, d_n^\nu\}=\eta^{\mu\nu}\delta_{m+n}
\end{equation}
The zero-frequency part of the Virasoro constraint provided the mass-shell condition
\begin{equation}
\label{eq2.12}
\alpha^{'}M^2=N+\frac{1}{2}
\end{equation} 
and the number operator is given by
\begin{equation}
\label{eq2.13}
N=N^\alpha+N^d
\end{equation} 
or
\begin{equation}
\label{eq2.14}
N=N^\alpha+N^b
\end{equation} 
where
\begin{equation}
\label{eq2.15}
N^\alpha=\sum_{m=1}^\infty \alpha_{-m}.\alpha_m
\end{equation} 
\begin{equation}
\label{eq2.16}
N^d=\sum_{m=1}^\infty md_{-m}.d_m
\end{equation} 
\begin{equation}
\label{eq2.17}
N^b=\sum\limits_{r=\frac {1} {2}} b_{-r}.b_r
\end{equation} 

The Virasoro operators for (NS) and (R) sectors respectively in terms of the oscillators are:
\begin{equation}
\label{eq2.18}
L_m=L_m^\alpha+L_m^b
\end{equation}
\begin{equation}
\label{eq2.19}
L_m=L_m^\alpha+L_m^d
\end{equation}
where $L_m^\alpha$ is the contribution from the bosonic modes and is expressed as

\begin{equation}
\label{eq2.20}
L_m^\alpha=\frac{1}{2}\sum_{n=-\infty}^{\infty}:\alpha_{-n}.\alpha_{m+n}:
\end{equation}
and the corresponding contribution of the fermionic modes (NS) sector is given as 
\begin{equation}
\label{eq2.21}
L_m^b=\frac{1}{2}\sum_{n=-\infty}^{\infty}(r+\frac{1}{2}m):b_{-r}.b_{m+r}:
\end{equation}
While the fermionic contribution in R sector is expressed as
\begin{equation}
\label{eq2.22}
L_m^d=\frac{1}{2}\sum_{n=-\infty}^{\infty}(n+\frac{1}{2}m):d_{-n}.b_{m+n}:
\end{equation}
For the fermionic generators or the modes of supercurrents one can write for (NS) and (R) sectors respectively as
\begin{equation}
\label{eq2.23}
G_r=\sum_{n=-\infty}^{\infty}:\alpha_{-n}.b_{r+n}:
\end{equation}
\begin{equation}
\label{eq2.24}
F_m=\sum_{n=-\infty}^{\infty}:\alpha_{-n}.d_{m+n}:
\end{equation}
When quantizing the RNS string the only requirement is that the positive modes of the Virasoro generators annihilate the physical state. For a physical state  $|\phi>$ in (NS) sector the super Virasoro constraints or the physical state conditions are
\begin{equation} 
\label{eq2.25}
G_r|\phi>=0, r>0
\end{equation}
\begin{equation}
\label{eq2.26}
L_n|\phi>=0, n>0
\end{equation}
\begin{equation}
\label{eq2.27}
(L_0-a_{NR}|\phi>)=0
\end{equation}
Similarly in R sector the physical-state conditions are
\begin{equation} 
\label{eq2.28}
F_n|\phi>=0, r>0
\end{equation}
\begin{equation}
\label{eq2.29}
L_n|\phi>=0, n>0
\end{equation}
\begin{equation}
\label{eq2.30}
(L_0-a_R|\phi>)=0
\end{equation}
where $a_{NR}=\frac{1}{2}$ and $a_R=0$ are the constants introduced for a normal-ordering ambiguity. These values have been determined to keep the negative norm states absent. \cite{bbs}
Light cone gauge an outcome of the reparameterization invariance of the bosonic string, is applied in RNS model of the superstring theory in an analogous way. The gauge choice $X^{+}(\sigma,\tau)=x^{+}+p^{+}\tau$, sufficient to gauge away the $+$components of all the nonzero mode oscillators in bosonic string theory, is supplemented in superstring theory by an additional freedom of applying local supersymmetric transformation $\psi^{+}=0$ without altering the existing gauge choice.\cite{gsw}

\setcounter{equation}{0}

\section{Coherent states for open NS superstring}

The bosonic part of the supercoherent state is defined in the same way as for the bosonic string \cite{hm}:
\begin{equation}
\label{eq3.1}
|\alpha_B,p>=\prod\limits_{n=1}^\infty\prod\limits_{j=1}^{9}\otimes|\alpha_{nB}^j>\otimes|\alpha_{nB}^0>\otimes|p>
\end{equation}
Where:
\begin{equation}
\label{eq3.2}    
|\alpha_{nB}^j>=\left(\frac {1} {\pi}\right)^{\frac {1} {4}}
e^{-\frac {(\alpha_{nB}^j)^2} {2}} e^{-\frac {|\alpha_{nB}^j|^2} {2}}
\int e^{-\frac {(y_n^j)^2} {2}} e^{  \sqrt{2}\alpha_{nB}^j y_n^j}|y_n^j>dy_n^j
\end{equation}
O, equivalently:
\begin{equation}
\label{eq3.3} 
|\alpha_{nB}^j>=e^{-\frac {|\alpha_{nB}^j|^2} {2}}\sum\limits_{m=0}^\infty\frac {\left(\alpha_{nB}^j\right)^m} {\sqrt{m!}}|m,j,n,B>
\end{equation}
The annihilation and creation operators are then defined as:
\begin{equation}
\label{eq3.4}
\hat{a}_{nB}^j=\frac {1} {\sqrt{n}}\hat{\alpha}_{nB}^j\;\;\;n>0
\end{equation}
\begin{equation}
\label{eq3.5}
\hat{a}^{+j}_{nB}=\frac {1} {\sqrt{n}}\hat{\alpha}_{-nB}^j\;\;\;n>0
\end{equation}
The action of the annihilation operators on the states $|m,j,n,B> $ is, as usual:
\begin{equation}
\label{eq3.6}
\hat{a}_{nB}^j|m,j,n,B>=\sqrt{m}|m-1,j.n,B>
\end{equation}
\begin{equation}
\label{eq3.7}
\hat{a}^{+j}_{nB}|m,j,n,B>=\sqrt{m+1}|m+1,j.n.B>
\end{equation}
As a consequence, the rule of the coherent state for a given $n$ then becomes:
\begin{equation}
\label{eq3.8}    
|||\alpha_{nB}^j>||=1
\end{equation}
For $\mu=0$ the annihilation and creation operators are defined as
\begin{equation}
\label{eq3.9}
\hat{a}_{nB}^0=\frac {1} {\sqrt{n}}\hat{\alpha}_{nB}^0\;\;\;n>0
\end{equation}
\begin{equation}
\label{eq3.10}
\hat{a}^{+0}_{nB}=\frac {1} {\sqrt{n}}\hat{\alpha}_{-nB}^0\;\;\;n>0
\end{equation}
According to the reference \cite{hm} the corresponding bosonic component of the supercoherent state is given by:
\begin{equation}
\label{eq3.11}    
|\alpha_{nB}^0>=\left(\frac {1} {\pi}\right)^{\frac {1} {4}}
e^{\frac {(\alpha_{nB}^0)^2} {2}} e^{\frac {|\alpha_{nB}^0|^2} {2}}
\int e^{\frac {(y_n^0)^2} {2}} e^{-\sqrt{2}\alpha_{nB}^0 y_n^0}|y_n^0>dy_n^0
\end{equation}
Its norm is null:
\begin{equation}
\label{eq3.12}    
|||\alpha_{nB}^0>||=0
\end{equation}
Since the coherent state has a null norm, that is
\begin{equation}
\label{eq3.13}
<\alpha_B,p^{'}|\alpha_B,p>=0
\end{equation}
It is then verified that
\begin{equation}
\label{eq3.14}
\int |\alpha_{nB}^0>\frac {d\alpha_{nB}^0} {\pi}<\alpha_{nB}^0|=0
\end{equation}
The fermionic component of the supercoherent state is defined as:
\begin{equation}
\label{eq3.15}
|\alpha_{rF}^\mu>=\frac {2\left(\eta^{\mu\mu}-\alpha_{rF}^\mu b_{r+\frac {1} {2}}^{\mu+}\right)}
{2+\eta^{\mu\mu}\overline{\alpha_{rF}^\mu}\alpha_{rF}^\mu}|0>
\end{equation}
where $\alpha_r^\mu$ is a Grassmann variable.
This state verifies:
\begin{equation}
\label{eq3.16}
b_{r+\frac {1} {2}}^{\mu+}|\alpha_{rF}^\mu>=\alpha_{rF}^\mu|\alpha_{rF}^\mu>
\end{equation}
The fermionic component
of the supercoherent state is then:
\begin{equation}
\label{eq3.17}
|\alpha_{F}>=\prod\limits_{r=0}^\infty\sum\limits_{\mu=0}^9|\alpha_{rF}^\mu>
\end{equation}
The corresponding standards turn out to be:
\begin{equation}
\label{eq3.18}
<\alpha_{rF}^\mu|\alpha_{rF}^\mu>=1
\end{equation}
\begin{equation}
\label{eq3.19}
<\alpha_{F}^\mu|\alpha_{F}^\mu>=1
\end{equation}
The supercoherent state is then defined as:
\begin{equation}
\label{eq3.20}
|\alpha>=|\alpha_B>\otimes|\alpha_F>
\end{equation}
This state verifies:
\begin{equation}
\label{eq3.21}
|||\alpha>||=0
\end{equation}
This is a consequence of: 
\begin{equation}
\label{eq3.22}
|||\alpha_B>||=0
\end{equation}
The annihilation and creation operators are defined as follows:
\begin{equation}
\label{eq3.23}
A_{nm}^{\mu\nu}=a_n^\mu\otimes b_{m+\frac {1} {2}}^\nu
\end{equation}
\begin{equation}
\label{eq3.24}
A_{nm}^{+\mu\nu}=a_n^{+\mu}\otimes b_{m+\frac {1} {2}}^{+\nu}
\end{equation}
The action of the annihilation operator on the supercoherent state is then:
\begin{equation}
\label{eq3.25}
A_{nm}^{\mu\nu}|\alpha>=\alpha_{nB}^\mu\alpha_{mF}^\nu|\alpha>
\end{equation}
Virasoro constraints have already been defined in, for example, the \cite{bbs} reference. For bosons we have:
\begin{equation}
\label{eq3.26}
L_0^{(\alpha)}=\frac {1} {2}\hat{\alpha}_0^2+\sum\limits_{s=1}^{\infty}
\hat{\alpha}_{-s}\cdot\hat{\alpha}_s
\end{equation}
\begin{equation}
\label{eq3.27}
L_k^{(\alpha)}=\frac {1} {2}\sum\limits_{s=-\infty}^{\infty}
:\hat{\alpha}_{k-s}\cdot\hat{\alpha}_s:\;\;\;\;;\;\;\;\;k\in\mathbb{Z}\;\;\;\;;\;\;\;\;k\neq 0
\end{equation}
For the fermions::
\begin{equation}
\label{eq3.28}
L_k^{(b)}=\frac {1} {2}\sum\limits_{r\in \mathbb{Z}+\frac {1} {2}}
\left(r+\frac {k} {2}\right):\hat{b}_{k+r}\cdot\hat{b}_{-r}:\;\;\;\;;\;\;\;\;k\in\mathbb{Z}
\end{equation}
Another constraint of Virasoro is also deduced from the supercurrent:
\begin{equation}
\label{eq3.29}
G_r=\sum\limits_{s=-\infty}^{\infty}
\hat{\alpha}_{-s}\cdot\hat{b}_{r+s}\;\;\;\;;\;\;\;\;r\in\mathbb{Z}+\frac {1} {2}
\end{equation}
The complete Virasoro constraint is:
\begin{equation}
\label{eq3.30}
L_k=L_k^{(\alpha)}+L_k^{(b)}
\end{equation}
The mass operator is defined as:
\begin{equation}
\label{eq3.31}
M^2=2\left[\sum\limits_{s=1}^{\infty}
\hat{\alpha}_{-s}\cdot\hat{\alpha}_s+
\sum\limits_{s=\frac {1} {2}}^\infty s\;
\hat{b}_{-s}\cdot\hat{b}_{s}-\frac {1} {2}\right]
\end{equation}
Virasoro constraints are satisfied on average by the supercoherent state. The result is:
\begin{equation}
\label{eq3.32}
<\alpha,p^{'}|\left(L_0-\frac {1} {2}\right)|\alpha,p>=0
\end{equation}
\begin{equation}
\label{eq3.33}
<\alpha,p^{'}|L_k|\alpha,p>=0\;\;\;\;;\;\;\;\;k>0
\end{equation}
\begin{equation}
\label{eq3.34}
<\alpha,p^{'}|G_r|\alpha,p>=0\;\;\;\;;\;\;\;\;r>0
\end{equation}
The mass of the coherent state turns out to be:
\begin{equation}
\label{eq3.35}
M^2=0
\end{equation}
Due to the equations (\ref{eq3.12}) and (\ref{eq3.12}) we can
remove the corresponding state for $\mu=0$ for bosons and redefine the supercoherent state as:
\begin{equation}
\label{eq3.36}
|\alpha_B,p>'=\prod\limits_{n=1}^\infty\prod\limits_{j=1}^{9}\otimes|\alpha_{nB}^j>\otimes|p>
\end{equation}
\begin{equation}
\label{eq3.37}
|\alpha>'=|\alpha_B>'\otimes|\alpha_F>
\end{equation}
It should then be verified:
\['<\alpha,p^{'}|\left(L_0-\frac {1} {2}\right)|\alpha,p>'=\]
\begin{equation} 
\label{eq3.38}
\left[\frac {p^2} {2}-\frac {1} {2}+\sum\limits_{n=1}^\infty n|\alpha_{nB}|^2
+\left(n+\frac {1} {2}\right) |\alpha_{nF}|^2\right]
\delta(p-p^{'})=0
\end{equation} 
and as a consequence:
\begin{equation} 
\label{eq3.39}
\frac {p^2} {2}-\frac {1} {2}+\sum\limits_{n=1}^\infty n|\alpha_{nB}|^2+\left(n+\frac {1} {2}\right) |\alpha_{nF}|^2
\end{equation}  
When $k>0$ they must satisfy:
\[ '<\alpha,p^{'}|L_k|\alpha,p>'='<\alpha,p^{'}|\left[
\sum\limits_{n=0}^k\sqrt{k-n}\sqrt{n}\hat{a}_n\hat{a}_{k-n}+\right.\]
\[+\sum\limits_{n=k+1}^\infty\sqrt{n-k}\sqrt{n}\hat{a}_n\hat{a}_{n-k}^+
\sum\limits_{n=1}^\infty \sqrt{n+k}\sqrt{n}\hat{a}_{k+n}\hat{a}_{n}+\]
\begin{equation}
\label{eq3.40}
\left.\sum\limits_{r=\frac {1} {2}}^\infty\left(\frac {k} {2}+r\right):b_{-r}b_{k+r}+
\sum\limits_{r=\frac {1} {2}}^\infty\left(\frac {k} {2}-r\right):b_{r}b_{k-r}\right]|\alpha,p>'=0
\end{equation}
The mass of the coherent state is now:
\begin{equation}
\label{eq3.41}
M^2\delta(p^{'}-p)=<\alpha,p^{'}|(2L_0-1-\alpha_0^2)|\alpha,p>
\end{equation}
And then:
\begin{equation}
\label{eq3.42}
M^2=2\sum\limits_{n=1}^{\infty}n|\alpha_n|^2+\left(n+\frac {1} {2}\right) |\alpha_{nF}|^2 -1
\end{equation}
{\bf Looking closely at (\ref{eq3.42}) we can see that by suitably choosing the $\alpha_n$ we can obtain states for which $M^2<0$ and $ <\alpha,p^{'}|\alpha,p>=\delta(p^{'}-p)$ that is, a tachyonic state whose norm is positive. For example if we select
$\alpha_{nB}=0$ and $\alpha_{nF}=0$.}

\nd
In the section 5 of the paper, we will prove that these results are similar with those obtained by defining the coherent states in the Light cone quantization formalism.

\setcounter{equation}{0}

\section{Mandel Parameter for supercoherent states of the NS superstring}

The definition of Mandel parameter \cite{man} $Q$ for a single mode of the string is given as 
$Q=Q_1+Q_2+Q_3+Q_4$ where:
\begin{equation}
\label{eq4.1} 
Q_1=-\frac{<\hat{A}^{+ij}_{mn}\hat{A}_{mn}^{ij}>-<(\hat{A}^{+ij}_{mn}\hat{A}_{mn}^{ij}\hat{A}^{+ij}_{mn}\hat{A}_{mn}^{ij})>+<\hat{A}^{+ij}_{mn}\hat{A}_{mn}^{ij}>^2}{<\hat{A}^{+ij}_{mn}\hat{A}_{mn}^{ij}>}
\end{equation}
And:
\begin{equation}
\label{eq4.2}
A_{nm}^{ij}=a_n^i\otimes b_{m+\frac {1} {2}}^j
\end{equation}
\begin{equation}
\label{eq4.3}
A_{nm}^{+ij}=a_n^{+i}\otimes b_{m+\frac {1} {2}}^{+j}
\end{equation}
For the redefined supercoherente state:
\begin{equation}
\label{eq4.4}
|\alpha>'=|\alpha_B>'\otimes|\alpha_F>
\end{equation}
We have then:
\begin{equation}
\label{eq4.5} 
<\hat{A}^{+ij}_{mn}\hat{A}_{mn}^{ij}>='<\alpha|\hat{A}^{+ij}_{mn}\hat{A}_{mn}^{ij}|\alpha>'
\end{equation}
And:
\begin{equation}
\label{eq4.6}
A_{nm}^{ij}|\alpha>=\alpha_{nB}^i\alpha_{mF}^j|\alpha>
\end{equation}
\begin{equation}
\label{eq4.7}
<\alpha|A_{nm}^{+ij}=<\alpha^*|\alpha_{nB}^{*i}\alpha_{mF}^{j}
\end{equation}
As a consequence:
\begin{equation}
\label{eq4.8} 
<\hat{A}^{+ij}_{mn}\hat{A}_{mn}^{ij}>=<\alpha_{nB}^{i}|\hat{a}^{+i}_{n}\hat{a}_{n}^{i}|\alpha_{nB}^{i}><\alpha_{mF}^{j}|\hat{b}^{+j}_{m+1/2}\hat{b}_{m+1/2}^{j}|\alpha_{mF}^{j}>
\end{equation}
The result is:
\begin{equation}
\label{eq4.9} 
<\hat{A}^{+ij}_{mn}\hat{A}_{mn}^{ij}>=|\alpha_{nB}^i|^2|\alpha_{mF}^j|^2
\end{equation}
Thus substituting the results in the equation for $Q$ parameter we could get the Mandel parameter as:
\begin{equation}
\label{eq4.10} 
Q_1=-\frac {|\alpha_{nB}^i|^2|\alpha_{mF}^j|^2-|\alpha_{nB}^i|^2|\alpha_{mF}^j|^2<\hat{A}^{ij}_{mn}\hat{A}_{mn}^{+ij}>+|\alpha_{nB}^i|^4|\alpha_{mF}^j|^4}{|\alpha_{nB}^i|^2|\alpha_{mF}^j|^2}
\end{equation}
With 
\begin{equation}
\label{eq4.11} 
'<\alpha|\hat{A}^{ij}_{mn}\hat{A}_{mn}^{+ij}|\alpha>'=<\alpha_{nB}^{i}|\hat{a}^{i}_{n}\hat{a}_{n}^{+i}|\alpha_{nB}^{i}><\alpha_{mF}^{j}|\hat{b}^{j}_{m+1/2}\hat{b}_{m+1/2}^{+j}|\alpha_{mF}^{j}>
\end{equation}
 Using commutation relation $[\hat{a}^{i}_{n},\hat{a}_{n}^{+i}]=1$,
$\hat{a}^{i}_{n}\hat{a}_{n}^{+i}=1+\hat{a}^{+i}_{n}\hat{a}_{n}^{i}$ and anticommutation relation ${\hat{b}^{j}_{m+1/2},\hat{b}_{m+1/2}^{+j}}$, $\hat{b}^{j}_{m+1/2}\hat{b}_{m+1/2}^{+j}=1-\hat{b}^{+j}_{m+1/2}\hat{b}_{m+1/2}^{j}$
we get
\begin{equation}
\label{eq4.12} 
'<\alpha|\hat{A}^{ij}_{mn}\hat{A}_{mn}^{+ij}|\alpha>'=<\alpha_{nB}^{i}|1+\hat{a}^{+i}_{n}\hat{a}_{n}^{i}|\alpha_{nB}^{i}><\alpha_{mF}^{j}|1-\hat{b}^{+j}_{m+1/2}\hat{b}_{m+1/2}^{j}|\alpha_{mF}^{j}>
\end{equation}
\begin{equation}
\label{eq4.13} 
<\hat{A}^{ij}_{mn}\hat{A}_{mn}^{+ij}>=\left(1+|\alpha_{nB}^{i}|^2\right)\left(1-|\alpha_{mF}^{j}|^2\right)
\end{equation}
\begin{equation}
\label{eq4.14} 
<\hat{A}^{ij}_{mn}\hat{A}_{mn}^{+ij}>=\left(1+|\alpha_{nB}^{i}|^2-|\alpha_{mF}^{j}|^2-|\alpha_{nB}^{i}|^2|\alpha_{mF}^{j}|^2\right)
\end{equation}
We obtain
\begin{equation}
\label{eq4.15} 
Q_1=|\alpha_{nB}^{i}|^2-|\alpha_{mF}^{j}|^2-2|\alpha_{nB}^{i}|^2|\alpha_{mF}^{j}|^2
\end{equation}
,for $\mu=i$, $\nu=0$ we have:
\begin{equation}
\label{eq4.16} 
Q_2=-\frac{<\hat{A}^{+i0}_{mn}\hat{A}_{mn}^{i0}>-<(\hat{A}^{+i0}_{mn}\hat{A}_{mn}^{i0}\hat{A}^{+i0}_{mn}\hat{A}_{mn}^{i0})>+<\hat{A}^{+i0}_{mn}\hat{A}_{mn}^{i0}>^2}{<\hat{A}^{+i0}_{mn}\hat{A}_{mn}^{i0}>}
\end{equation}
\begin{equation}
\label{eq4.17} 
'<\alpha|\hat{A}^{+i0}_{mn}\hat{A}_{mn}^{i0}|\alpha>'=<\alpha_{nB}^{i}|\hat{a}^{+i}_{n}\hat{a}_{n}^{i}|\alpha_{nB}^{i}><\alpha_{mF}^{0}|\hat{b}^{+0}_{m+1/2}\hat{b}_{m+1/2}^{0}|\alpha_{mF}^{0}>
\end{equation}
\begin{equation}
\label{eq4.18} 
<\hat{A}^{+i0}_{mn}\hat{A}_{mn}^{i0}>=|\alpha_{nB}^i|^2|\alpha_{mF}^0|^2
\end{equation}
Using commutation relation $[\hat{a}^{i}_{n},\hat{a}_{n}^{+i}]=1$,
$\hat{a}^{i}_{n}\hat{a}_{n}^{+i}=1+\hat{a}^{+i}_{n}\hat{a}_{n}^{i}$ and anticommutation relation $\{\hat{b}^{0}_{m+1/2},\hat{b}_{m+1/2}^{+0}\}=-1$, $\hat{b}^{0}_{m+1/2}\hat{b}_{m+1/2}^{+0}=-1-\hat{b}^{+0}_{m+1/2}\hat{b}_{m+1/2}^{0}$
we get:
\begin{equation}
\label{eq4.19} 
<\alpha|\hat{A}^{i0}_{mn}\hat{A}_{mn}^{+i0}|\alpha>=<\alpha_{nB}^{i}|1+\hat{a}^{+i}_{n}\hat{a}_{n}^{i}|\alpha_{nB}^{i}><\alpha_{mF}^{0}|-1-\hat{b}^{+0}_{m+1/2}\hat{b}_{m+1/2}^{0}|\alpha_{mF}^{0}>
\end{equation}
This is:
\begin{equation}
\label{eq4.20} 
<\hat{A}^{i0}_{mn}\hat{A}_{mn}^{+i0}>=\left(1+|\alpha_{nB}^{i}|^2\right)\left(-1-|\alpha_{mF}^{0}|^2\right)
\end{equation}
As a consequence:
\begin{equation}
\label{eq4.21} 
<\hat{A}^{i0}_{mn}\hat{A}_{mn}^{+i0}>=\left(-1-|\alpha_{nB}^{i}|^2-|\alpha_{mF}^{0}|^2-|\alpha_{nB}^{i}|^2|\alpha_{mF}^{0}|^2\right)
\end{equation}
The result for $Q_2$ is:
\begin{equation}
\label{eq4.22} 
Q_2=-\left(2+|\alpha_{nB}^{i}|^2+|\alpha_{mF}^{0}|^2+2|\alpha_{nB}^{0}|^2|\alpha_{mF}^{0}|^2\right)
\end{equation}
For $\mu=0$, $\nu=j$ we have:
\begin{equation}
\label{eq4.23} 
Q_3=-\frac{<\hat{A}^{+0j}_{mn}\hat{A}_{mn}^{0j}>-<(\hat{A}^{+0j}_{mn}\hat{A}_{mn}^{0j}\hat{A}^{+0j}_{mn}\hat{A}_{mn}^{0j})>+<\hat{A}^{+0j}_{mn}\hat{A}_{mn}^{0j}>^2}{<\hat{A}^{+0j}_{mn}\hat{A}_{mn}^{0j}>}
\end{equation}
We calculate the terms and use then for calculating $Q_3$ as
\begin{equation}
\label{eq4.24} 
<\alpha|\hat{A}^{+0j}_{mn}\hat{A}_{mn}^{0j}|\alpha>=<\alpha_{nB}^{0}|\hat{a}^{+0}_{n}\hat{a}_{n}^{0}|\alpha_{nB}^{0}><\alpha_{mF}^{j}|\hat{b}^{+j}_{m+1/2}\hat{b}_{m+1/2}^{j}|\alpha_{mF}^{j}>
\end{equation}
\begin{equation}
\label{eq4.25} 
<\hat{A}^{+0j}_{mn}\hat{A}_{mn}^{0j}>=|\alpha_{nB}^0|^2|\alpha_{mF}^j|^2
\end{equation}
Using commutation relation $[\hat{a}^{0}_{n},\hat{a}_{n}^{+0}]=-1$,
$\hat{a}^{0}_{n}\hat{a}_{n}^{+0}=-1+\hat{a}^{+0}_{n}\hat{a}_{n}^{0}$ and anticommutation relation $\{\hat{b}^{0}_{m+1/2},\hat{b}_{m+1/2}^{+0}\}=1$, $\hat{b}^{0}_{m+1/2}\hat{b}_{m+1/2}^{+0}=1-\hat{b}^{+0}_{m+1/2}\hat{b}_{m+1/2}^{0}$
\begin{equation}
\label{eq4.26} 
<\alpha|\hat{A}^{0j}_{mn}\hat{A}_{mn}^{+0j}|\alpha>=<\alpha_{nB}^{0}|-1+\hat{a}^{+0}_{n}\hat{a}_{n}^{0}|\alpha_{nB}^{0}><\alpha_{mF}^{j}|1-\hat{b}^{+j}_{m+1/2}\hat{b}_{m+1/2}^{j}|\alpha_{mF}^{j}>
\end{equation}
\begin{equation}
\label{eq4.27} 
<\hat{A}^{0j}_{mn}\hat{A}_{mn}^{+0j}>=\left(|\alpha_{nB}^{0}|^2\right)\left(1-|\alpha_{mF}^{j}|^2\right)
\end{equation}
since
\begin{equation}
\label{eq4.28}    
|||\alpha_{nB}^0>||=0
\end{equation}
\begin{equation}
<\hat{A}^{0j}_{mn}\hat{A}_{mn}^{+0j}>=\left(|\alpha_{nB}^{0}|^2-|\alpha_{nB}^{0}|^2|\alpha_{mF}^{j}|^2\right)
\end{equation}
Then we have:
\begin{equation}
\label{eq4.29} 
Q_3=-\left(1-|\alpha_{nB}^{0}|^2+2|\alpha_{nB}^{0}|^2|\alpha_{mF}^{j}|^2\right)
\end{equation}
Similarly,for $\mu=0$, $\nu=0$ we have:
\begin{equation}
\label{eq4.30} 
Q_4=-\frac{<\hat{A}^{+00}_{mn}\hat{A}_{mn}^{00}>-<(\hat{A}^{+00}_{mn}\hat{A}_{mn}^{00}\hat{A}^{+00}_{mn}\hat{A}_{mn}^{00})>+<\hat{A}^{+00}_{mn}\hat{A}_{mn}^{00}>^2}{<\hat{A}^{+00}_{mn}\hat{A}_{mn}^{00}>}
\end{equation}
Again we calculate the terms and use then for calculating $Q$ as
\begin{equation}
\label{eq4.31} 
<\alpha|\hat{A}^{+00}_{mn}\hat{A}_{mn}^{00}|\alpha>=<\alpha_{nB}^{0}|\hat{a}^{+0}_{n}\hat{a}_{n}^{0}|\alpha_{nB}^{0}><\alpha_{mF}^{0}|\hat{b}^{+0}_{m+1/2}\hat{b}_{m+1/2}^{0}|\alpha_{mF}^{0}>
\end{equation}
\begin{equation}
\label{eq4.32} 
<\hat{A}^{+00}_{mn}\hat{A}_{mn}^{00}>=|\alpha_{nB}^0|^2|\alpha_{mF}^0|^2
\end{equation}
Using commutation relation $[\hat{a}^{0}_{n},\hat{a}_{n}^{+0}]=-1$,
$\hat{a}^{0}_{n}\hat{a}_{n}^{+0}=-1+\hat{a}^{+0}_{n}\hat{a}_{n}^{0}$ and anticommutation relation $\{\hat{b}^{0}_{m+1/2},\hat{b}_{m+1/2}^{+0}\}=-1$, $\hat{b}^{0}_{m+1/2}\hat{b}_{m+1/2}^{+0}=-1-\hat{b}^{+0}_{m+1/2}\hat{b}_{m+1/2}^{0}$
\begin{equation}
\label{eq4.33} 
<\alpha|\hat{A}^{00}_{mn}\hat{A}_{mn}^{+00}|\alpha>=<\alpha_{nB}^{0}|-1+\hat{a}^{+0}_{n}\hat{a}_{n}^{0}|\alpha_{nB}^{0}><\alpha_{mF}^{0}|-1-\hat{b}^{+0}_{m+1/2}\hat{b}_{m+1/2}^{0}|\alpha_{mF}^{0}>
\end{equation}
\begin{equation}
\label{eq4.34} 
<\hat{A}^{00}_{mn}\hat{A}_{mn}^{+00}>=\left(|\alpha_{nB}^{0}|^2\right)\left(-1-|\alpha_{mF}^{0}|^2\right)
\end{equation}
since
\begin{equation}
\label{eq4.35}    
|||\alpha_{nB}^0>||=0
\end{equation}
\begin{equation}
\label{eq4.36} 
<\hat{A}^{00}_{mn}\hat{A}_{mn}^{+00}>=\left(-|\alpha_{nB}^{0}|^2-|\alpha_{nB}^{0}|^2|\alpha_{mF}^{0}|^2\right)
\end{equation}
Then we have:
\begin{equation}
\label{eq4.37} 
Q_4=-\left(1+|\alpha_{nB}^{0}|^2+2|\alpha_{nB}^{0}|^2|\alpha_{mF}^{0}|^2\right)
\end{equation}

\setcounter{equation}{0}
\section{Supercoherent States in the light cone quantization formalism}
Analytically not different than covariant formalism but with pedagogical supremacy, the light cone quantization approach has been successful in dealing some advanced problems. The specialty of the light cone quantization is that it preserves the physical degrees of freedom and succeeds in eliminating the part of the string degrees of freedom and satisfyingly fix the residual gauge by setting 
$X^{+}=x^{+}+p^{+}\tau$, $\psi^+=0$  and in consequence the Lorentz covariance is explicitly broken. 
The zero modes $x^+$ and $p^+$ play the role of the integration constants, chosen arbitrarily, and are left as free parameters. The advantage of this gauge choice is that it allows every point on the string to be at the same value of time. Classical interpretation of this gauge fixing implies to set the oscillator coefficients $\alpha_n^+$ to zero for $n\ne0$.
The mode expansion scheme of the oscillator is not disturbed, but the formalism scrapes an infinite set of modes to zero thus formulating and expressing everything in terms of the transverse oscillators alone.
The light cone quantization takes place in the same way as in canonical formalism but, for the transverse oscillators only. 
Operators $\alpha_n^-$ and $b_r^-$ are defined as
\begin{equation}
\label{5.1}
{\alpha}_n^-=\frac{1}{2p^{+}}\sum\limits_{j=2}^{10}\left[
\sum\limits_{k}:\alpha_{n-k}^j.\alpha_{k}^j:+
\sum\limits_{r\in\mathbb{Z}+\frac {1} {2}}\left(r-\frac {n} {2}\right):b_{n-r}^j.b_r^j:\right]
-a\delta_{n,0}
\end{equation}
\begin{equation}
\label{5.2}
b_r^-=\frac{1}{p^{+}}\sum\limits_{j=2}^{10}
\sum\limits_{s\in\mathbb{Z}+\frac {1} {2}}:\alpha_{r-s}^j.b_s^j:
\end{equation}
The square mass operator is then
\begin{equation}
\label{eq5.3}
M^2=2\sum\limits_{n=1}^{\infty}n|\alpha_{nB|}^2+\left(n+\frac {1} {2}\right) |\alpha_{nF}|^2 -1
\end{equation}
The mass-shell condition obtained is same in both the formalism ,With a lone difference that in light cone treatment the contribution comes from only the transverse oscillators. To preserve the Lorentz invariance of the theory, the parameter $a$ must be equal to $1/2$ and the dimension $D$ has to be $10$. The other advantageous aspect is that the light cone formalism is considered to be ghost free.\cite{gsw}
Thus the treatment of supercoherent states in the light cone quantization is similar to that of coherent states of open  superstring with the only difference that $j=2,3,4,...10$. Following the same mathematical procedure we express the supercoherent states for the open string in light cone treatment as
\begin{equation}
\label{eq5.4}
|\alpha_{F}>'=\prod\limits_{r=0}^\infty\sum\limits_{j=2}^9|\alpha_{rF}^j>
\end{equation}

\begin{equation}
\label{eq5.5}
|\alpha_B,p>'=\prod\limits_{n=1}^\infty\prod\limits_{j=2}^{9}\otimes|\alpha_{nB}^j>\otimes|p>
\end{equation}

\begin{equation}
\label{eq5.6}
|\alpha>'=|\alpha_B>'\otimes|\alpha_F>'
\end{equation}
Thus we have constructed the supercoherent states of open  superstring in light cone quantization formalism and interestingly found the similar results as that obtained from the old covariant formalism. The mass of the supercoherent state will also be same as that of the results obtained in covariant treatment.

\setcounter{equation}{0}

\section{Conclusion}

\nd 
In this paper we have obtained the Glauber supercoherent states for the NS superstring in the two instances, one using old covariant formalism and one in the light cone quantization scheme, followed in this paper.  Keeping intact the original definition of coherent states of harmonic oscillator we rigorously defined the supercoherent states for open superstring in the covariant quantization formalism. The supercoherent states thus obtained in this case, identically satisfied the super Virasoro constraints at mean value. The point which invokes interest is that the supercoherent states have zero mass very similar to that of the Glauber's coherent states for the electromagnetic fields \cite{glau}, which substantiates the covariant quantization approach followed by us in establishing the supercoherent states of the superstring. The fact that the identity resolution for the temporal bosonic components of the supercoherent state is null enables us to successfully redefine them into the  supercoherent states of non-zero mass.\\
The behaviour and the properties of the supercoherent states obtained by light cone quantization approach is same as that of the redefined states obtained using the covariant formalism, thus further strengthens the validity of both the approaches followed in the paper. 
 The evaluation of the Mandel parameter  for the supercoherent states to see the statistical nature of their probability distributions became the spontaneous extension of the paper. The results were exceptionally appealing for being sub-Poissonian, Poissonian and super-Poissonian depending on the value of the parameter being negative, null and positive respectively.
\nd {\bf An important fact we found is as :
Looking closely for example at (\ref{eq3.42}), we can see and infer that by a suitable choice of the $\alpha_n$ (say for example $\alpha_{nB}=\alpha_{nF}=0$ ), we come across with positive norm states bearing imaginary mass, i.e the states with $ <\alpha,p^{'}|\alpha,p>=\delta(p^{'}-p)$ and $M^2<0$ which conclusively  corresponds to a tachyonic state with positive norm.}

\end{document}